\documentstyle[psfig]{mn}
\newcommand{\arcm}{$^{\prime}$}
\newcommand{\arcs}{$^{\prime\prime}$}
\newcommand{\arcsd}[2]{$#1^{\prime\prime}\!\!.#2$}
\newcommand{\ra}[4]{$#1^{\mathrm{h}}#2^{\mathrm{m}}#3^{\mathrm{s}}\!\!.#4$}
\newcommand{\dec}[3]{$#1^{\circ}#2^{\prime}#3^{\prime\prime}$}
\newcommand{\magn}[2]{$#1^{\mathrm{m}}\!\!.#2$}

\title[NGC~6251 and its companion galaxies]
      {A spectroscopic study of NGC~6251 and its companion galaxies}

\author[P. N. Werner, D. M. Worrall and M. Birkinshaw]
       {P. N. Werner, D. M. Worrall and M. Birkinshaw \\
Department of Physics, University of Bristol, Tyndall Avenue, Bristol BS8 1TL}

\date{3 May 2000}  
\pubyear{2000}
\begin{document}
\maketitle

\begin{abstract}

Measurements of the velocities of galaxies thought to be associated
with the giant radio galaxy NGC~6251 confirm the presence of a poor
cluster with a systemic redshift of $\overline{z}= 0.0244 \pm 0.0004$
and a line-of-sight velocity dispersion of $\sigma_{z}= 283\, (+109,
-52)$ km~s$^{-1}$.  This suggests a cluster atmosphere temperature of
$T = 0.7\, (+0.6, -0.2)\,\rm{keV}$, which is not enough to confine the
radio jet by gas pressure.  The core of NGC~6251 shows strong emission
lines of [O~III] and H$\alpha$~+~[N~II], but there is no evidence for
line emission from the jet (detected in optical continuum by Keel
(1988)).

\end{abstract}


\section{Introduction}\label{sec:intro}

NGC~6251 is an m$_{B}$ = \magn{13}{6} (de Vaucouleurs et al.\,1991)
giant E2 radio galaxy known for its remarkable radio jet, which is
aligned (within a few degrees) from pc to Mpc scales (see Readhead,
Cohen \&\ Blandford 1978, Perley, Bridle \&\ Willis 1984 and Jones et
al.\,1986 for pc-scale observations, kpc-scale observations, and both,
respectively).  Keel (1988) detected the radio jet in optical
continuum emission and found that the spectrum was unusually steep.
Arguments that the galaxy contains a supermassive black hole with mass
$\sim 10^{9} M_{\odot}$ (Young et al.\,1979) were recently supported
by HST observations of the nuclear gas and dust disc on a scale of a
few $\times100$ pc (Ferrarese \&\ Ford 1999).  NGC~6251 has been
described as relatively isolated, but loosely associated with the
Zwicky cluster Zw~1609.0+8212.

This cluster, which Zwicky (1968) described as `medium compact' and
`near' with a population of 260, was found by Gregory \&\ Connolly
(1973) to have two main concentrations.  One of them is A2247, listed
as a separate cluster by Abell (1958).  Like NGC~6251, it lies on the
Eastern boundary of Zw~1609.0+8212.  Its systemic velocity is 11670
km~s$^{-1}$ (Sargent, Readhead \&\ de Bruyn 1982) and its interest
lies mainly in the fact that it contains a striking chain of
galaxies.  The Western part of the Zwicky cluster surrounds IC~1143.
The three galaxies which Gregory \&\ Connolly (1973) ascribed to this
cluster segment have velocities close to 6500 km~s$^{-1}$.  NGC~6251
appears to depart from this velocity by $\sim 1000$ km~s$^{-1}$, and
Prestage \&\ Peacock (1988) concluded that the local galaxy density is
so low that any atmosphere would more likely be associated with the
galaxy than the cluster.

Birkinshaw \&\ Worrall (1993) observed NGC~6251 with the \it ROSAT \rm
PSPC and detected a spatially unresolved X-ray concentration,
containing about 90\% of the 0.1--2 keV flux and coincident with the
radio core.  The remaining $\sim$10\% of the flux appeared to
originate from an extended emission component, probably a gaseous
atmosphere with a scale of $\sim$3\arcm\ (130 kpc)\footnote{In this paper we
adopt a Hubble constant $H_{0}$ = 50 km~s$^{-1}$ Mpc$^{-1}$.} FWHM.
They obtained a satisfactory fit to the PSPC radial profile by
superposing a point source and an isothermal $\beta$-model.  The
limited number of counts from the modelled extended emission did not
allow a useful gas temperature measurement to be made.  However,
Birkinshaw \&\ Worrall showed that for the outer parts of the
kpc-scale radio jet to be confined by atmospheric pressure, a gas
temperature of $T \geq 5$ keV is required.  Feature A (the brightest
part of the jet, only 10\arcs -- 40\arcs\ or $\sim$ 7 -- 29 kpc from the
nucleus) can be confined by a gas temperature of $T \geq 2$ keV.  The
galaxy's internal stellar velocity dispersion of $\sigma_{z}= 293 \pm
20$ km~s$^{-1}$ measured by Heckman, Carty \&\ Bothun (1985a) suggests
that the stellar gravitational field cannot hold an atmosphere this
hot.  In this paper, we demonstrate the existence of a cluster
associated more strongly with NGC~6251 than Zw~1609.0+8212 based on
velocities for 13 galaxies including IC~1143, and we use the velocity
dispersion of this cluster to estimate the temperature of the cluster
atmosphere. We discuss the consequences for pressure confinement of
NGC~6251's jet.

\section{Observations and data reduction}\label{sec:obs}

\begin{table*}
\caption{Observational details for the low-resolution spectra}
\label{tab:obsl}
\begin{tabular}{llcccc}
\hline
Object & Name       &\multicolumn{2}{c}{Coordinates (2000)} & Exp.(s) & Date observed  \\ 
       &            &       R.A.         &        Dec.      &         &                \\
\hline
1      & NGC 6251   & \ra{16}{32}{34}{8} & \dec{82}{32}{17} & 1800    & 22 Sept.\ 1992 \\
2      & ---        & \ra{16}{32}{47}{7} & \dec{82}{33}{25} & 900     & $\prime\prime$ \\
3      & NGC 6252   & \ra{16}{32}{43}{3} & \dec{82}{34}{36} & 600     & $\prime\prime$ \\
4      & Zw 367-010 & \ra{16}{22}{46}{1} & \dec{82}{23}{45} & 900     & 23 Sept.\ 1992 \\
5      & UGC 10581  & \ra{16}{43}{11}{6} & \dec{82}{37}{51} & 450     & $\prime\prime$ \\
6a     & UGC 10464a & \ra{16}{28}{34}{6} & \dec{82}{53}{16} & 450     & $\prime\prime$ \\
6b     & UGC 10464b & $\prime\prime$     & $\prime\prime$   & $\prime\prime$ & $\prime\prime$ \\
7      & Zw 367-015 & \ra{16}{37}{05}{5} & \dec{82}{00}{37} & 600     & $\prime\prime$ \\
8      & UGC 10431  & \ra{16}{25}{18}{3} & \dec{81}{47}{57} & 600     & $\prime\prime$ \\
9      & UGC 10604  & \ra{16}{47}{50}{2} & \dec{81}{50}{50} & 600     & $\prime\prime$ \\
10     & Zw 367-016 & \ra{16}{38}{48}{5} & \dec{81}{41}{28} & 600     & $\prime\prime$ \\
11a    & UGC 10471a & \ra{16}{30}{28}{5} & \dec{81}{33}{44} & 600     & $\prime\prime$ \\
11b    & UGC 10471b & $\prime\prime$     & $\prime\prime$   & $\prime\prime$ & $\prime\prime$ \\
12     & IC 1143    & \ra{15}{30}{48}{5} & \dec{82}{27}{34} & 300     & $\prime\prime$ \\
13     & IC 1139    & \ra{15}{29}{18}{3} & \dec{82}{35}{15} & 300     & $\prime\prime$ \\
\hline
\end{tabular}
\end{table*}

\begin{table*}
\caption{Observational details for the high-resolution spectra of NGC~6251}
\label{tab:obsh}
\begin{tabular}{ccccl}
\hline
Slit Position         & $\lambda$ range & Exp. time  & Date observed  & Comments     \\ 
Angle ($^{\circ}$)& (\AA)       & (s)        &                &                      \\
\hline
115.3            & 4874--5526      & 1800       & 22 Sept.\ 1992 & along jet         \\
\multicolumn{4}{c}{}                                    & H$\beta$,[O III], Mg$_{b}$ \\
115.3            & 6274--6926      & 1800       & $\prime\prime$ & along jet         \\
\multicolumn{4}{c}{}                                             & H$\alpha$, [N II] \\
25.3             & 4874--5526      & 1800       & 23 Sept.\ 1992 & perpendicular to jet \\
\multicolumn{4}{c}{}                                    & H$\beta$,[O III], Mg$_{b}$ \\
25.3             & 6274--6926      & 1800       & $\prime\prime$ & perpendicular to jet \\
\multicolumn{4}{c}{}                                             & H$\alpha$, [N II] \\
\hline
\end{tabular}
\end{table*}

Observations of the galaxies listed in Table~\ref{tab:obsl} were
carried out on the nights of 1992 September 22 and 23, using the Red
Channel spectrograph on the Multiple Mirror Telescope (MMT).  All
images were obtained on a $800\times800$ CCD (binned 2/1) with $(15
\mu \rm m)^{2}$ pixels using a long slit of dimensions
180\arcs$\times$\arcsd{1}{5}.  Low-resolution spectra were taken with
a 300 line mm$^{-1}$ grating, which had a useable range of 2560 \AA\
centred on $\sim$ 6000 \AA\, and had spatial and wavelength
dispersions of \arcsd{0}{5}\ and 3.2 \AA\ per pixel, respectively.  A
resolution of $\sim$~10 \AA\ (FWHM) was obtained in this instrument
configuration.

The galaxies listed in Table~\ref{tab:obsl} are numbered in order of
increasing distance from NGC~6251 (which is object 1).  Our sample
(which is not complete or magnitude-limited) includes two galaxies
with double cores, objects 6 and 11.  Although in both cases the two
nuclei are listed separately (and labelled \emph{a} and \emph{b}), the
average redshifts were used as single measurements in subsequent
cluster-related calculations, as they are clearly bound systems.

In addition, high-resolution spectra of NGC~6251 were obtained (see
Table~\ref{tab:obsh}) with a 1200 line mm$^{-1}$ grating, resulting in
a wavelength range of 652 \AA\ and a dispersion of 0.9 \AA\ per pixel.
Spectra centred on 5200 \AA\ and 6600 \AA\ were obtained along the
radio jet (at a position angle of $115.3^{\circ}$) and perpendicular
to it (P.A. = $25.3^{\circ}$).  The resolution of these spectra was
2--3~\AA\ (FWHM).  The objective of these observations was to search
for emission lines in the optical knots detected in continuum by Keel
(1988), and measure the rotation velocity and direction for the stars
and gas in the galaxy.

The reduction and analysis of our data were carried out using the
Image Reduction and Analysis Facility (IRAF).  The CCD images were
de-biased, trimmed and flat-fielded using the CCDPROC task.  Cosmic
ray hits and bad lines or columns on the CCD were removed by
interpolating across the smallest dimension of the affected regions
with the FIXPIX routine.  For the low-resolution case, where only a
nuclear one-dimensional spectrum was required, the spectra were
sky-subtracted using the BACKGROUND task, then wavelength-calibrated
using HeNeAr comparison spectra, traced and extracted with the
APEXTRACT package.

The data reduction in the high-resolution case was somewhat
complicated by the fact that information in the spatial direction was
to be preserved, so that the core spectrum could not simply be traced
along the geometrically distorted two-dimensional image.  A good
alignment (subsequently found to be accurate to typically a few tenths
of a pixel) of the dispersion axis and spatial axis with the CCD's
columns and lines, respectively, was carried out using the LONGSLIT
package.  HeNeAr calibration lines and the trace of the galaxy nucleus
were used to fit a two-dimensional geometrical rectification function
to the data.  This correction was then applied to the image, and 1-D
spectra at varying offsets from the core of the galaxy were
subsequently extracted.

\begin{figure}
\psfig{file=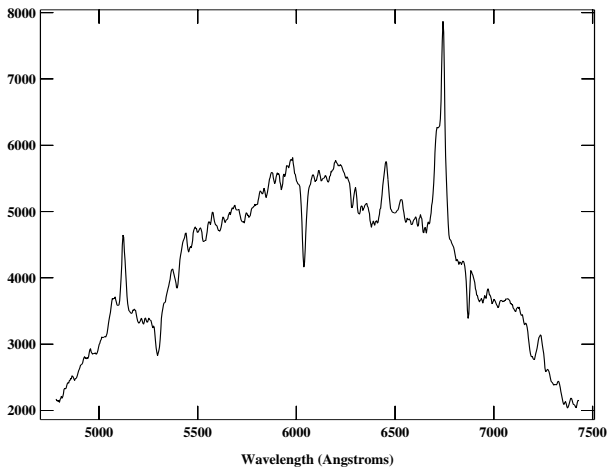,angle=270,width=9.5cm}

\caption{Raw (un-fluxed) spectrum (in MMT counts) of the core of
NGC~6251 in the observed frame.  Note the strong [O III], [O I] and
H$\alpha$ + [N II] lines, as well as stellar absorption features.  The
resolution is $\sim$10 \AA\ (FWHM).}

\label{fig:corespec}
\end{figure}

Figure~\ref{fig:corespec} is an un-fluxed spectrum of the core of
NGC~6251 from our low-resolution data.

\section{Redshifts}\label{sec:redshifts}

\subsection{Measurements}

Seven of the fifteen spectra showed emission lines which enabled
redshifts to be determined.  Table~\ref{tab:vem} shows which lines
were used and the resultant heliocentric velocities.  The error bars
take account of the scatter of the velocities calculated from the
various lines, as well as the errors on the individual line
wavelengths.  Due to a slight defocussing effect which is particularly
noticeable in the emission lines, an error of 10\% of the FWHM of the
line is included in these latter errors.

We measure the absorption redshifts of the galaxies by the standard
method of cross-correlating their stellar spectra with the stellar
absorption-line spectrum of an object of known redshift (Tonry \&
Davis 1979).  The cross-correlation was carried out over the
wavelength range 5000--5450 \AA\ and 5950--6200 \AA\ (including the
Mg$_{b}$ and Na D features while avoiding poorly subtracted night sky
lines around 5600 \AA) using the IRAF task FXCOR in the RV package.

The template used for the cross-correlation was a very high
signal-to-noise spectrum of NGC~7619 obtained from a 600s exposure
(with the 300 line mm$^{-1}$ grating) during the 1992 September 22
observing run.  NGC~7619 is a \magn{12}{1} E2 galaxy with a velocity
of $V=3747 \pm 20$ km s$^{-1}$ (Huchra et al.\,1983).  More recent
velocities for NGC~7619 (e.g. de Vaucouleurs et al.\,1991, Huchra et
al.\,1995) are in reasonable agreement, so that any systematic error
introduced into our results by the choice of template velocity is
likely to be less than a few tens of km s$^{-1}$.

The cross-correlation velocities of the galaxies in our sample are
shown in Table~\ref{tab:vabs}.

\begin{table}
\caption{Redshifts obtained from emission lines.}
\label{tab:vem}
\begin{tabular}{llr@{\,$\pm$\,}l} \hline
Object  &  Emission lines used           & \multicolumn{2}{c}{$V^{\rm(em)}$}   \\ 
        &                                & \multicolumn{2}{c}{(km s$^{-1}$)}   \\
\hline
1       & [O III]\ 4958, [O III]\ 5007, [O I]\ 6300     & 7276  & 75           \\
        & [N II] 6584                                   & \multicolumn{2}{c}{} \\
4       & [O III]\ 4958, [O III]\ 5007, [N II] 6563	& 6930  & 75           \\
        & H$\alpha$, [N II] 6584			& \multicolumn{2}{c}{} \\
6a	& [O III]\ 4958, [O III]\ 5007, [N II] 6563	& 7723  & 47           \\
	& H$\alpha$, [N II] 6584			& \multicolumn{2}{c}{} \\
7	& [O III]\ 4958, [O III]\ 5007, H$\alpha$, 	& 4141  & 41           \\
	& [N II] 6584, [S II] 6716, [S II] 6731		& \multicolumn{2}{c}{} \\
8	& [N II] 6563, H$\alpha$, [N II] 6584		& 7635  & 63           \\
%9	& H$\alpha$, [N II] 6584			& 7261  & 73          \\
10	& H$\alpha$, [N II] 6584			& 16648 & 51            \\
11a	& H$\beta$, H$\alpha$, [N II] 6584		& 7128  & 54           \\
\hline
\end{tabular}
\end{table}

\begin{table*}
\caption{Results of the cross-correlations and peculiar velocities relative to the cluster's systemic velocity}
\label{tab:vabs}
\begin{tabular}{lr@{\,$\pm$\,}lr@{\,$\pm$\,}lrlrlcc}
\hline
Object & \multicolumn{2}{c}{$V^{\rm(abs)}$} & \multicolumn{2}{c}{$z^{\rm(abs)}$} & \multicolumn{1}{c}{$V_{\rm pec}$}& &\multicolumn{1}{c}{$V_{\rm pec}/\sigma_{z}$} & & \multicolumn{2}{c}{Distance from NGC 6251} \\
    & \multicolumn{2}{c}{(km~s$^{-1}$)} & \multicolumn{2}{c}{} & \multicolumn{1}{c}{(km~s$^{-1}$)} & & & & (arcmin) & (Mpc)  \\ \hline
1   & 7459  & 22           &  0.0249 & 0.0001       &   260     &     &   0.67   &        & 0    & 0 \\
2   & 6913  & 56           &  0.0231 & 0.0002       & $-273$    &     & $-0.70$  &        & 1.21 & 0.052 \\
3   & 7240  & 46           &  0.0242 & 0.0002       &    46     &     &   0.11   &        & 2.33 & 0.101 \\  
4   & 7001  & 38           &  0.0234 & 0.0001       & $-187$    &     & $-0.48$  &        & 21.1 & 0.911\\
5   & 7223  & 44           &  0.0241 & 0.0001       &    30     &     &   0.08   &        & 21.3 & 0.919 \\
6a  & 7797  & 43           &  0.0260 & 0.0001       &   567 &$^{\,a}$ &   1.46  &$^{\,a}$ & 22.3 & 0.964\\    
6b  & 7748  & 43           &  0.0258 & 0.0001       &$\prime\prime$&$^{\,a}$ &$\prime\prime$&$^{\,a}$ &$\prime\prime$&$\prime\prime$ \\
7   & 4224  & 140          &  0.0138 & 0.0004       &   --- &$^{\,b}$ &   ---   &$^{\,b}$ & 33.0 & 1.42 \\
8   &\multicolumn{2}{c}{--- $^{\,c}$} & \multicolumn{2}{c}{--- $^{\,c}$} & 432 &$^{\,d}$ & 1.11& $^{\,d}$ & 46.8 & 2.02 \\
9   & 7554  & 56           &  0.0252 & 0.0002       &   353     &     &   0.91   &        & 51.8 & 2.24     \\
10  &\multicolumn{2}{c}{--- $^{\,c}$} & \multicolumn{2}{c}{--- $^{\,c}$}& --- &$^{\,b,d}$ & ---&$^{\,b,d}$&52.4 &2.26 \\
11a & 7158  & 102          &  0.0239 & 0.0003       & $-36$ &$^{\,a}$ & $-0.09$ &$^{\,a}$ & 58.7 & 2.54 \\
11b & 7154  & 52           &  0.0239 & 0.0002       &$\prime\prime$ &$^{\,a}$ &$\prime\prime$ &$^{\,a}$ & $\prime\prime$ &$\prime\prime$  \\
12  & 6548  & 40           &  0.0218 & 0.0001       & $-629$    &     & $-1.62$ &         & 121 & 5.21 \\
13  & 6616  & 71           &  0.0221 & 0.0002       & $-563$    &     & $-1.45$ &         & 123 & 5.29\\
\hline
\end{tabular}

$^{a}$\ average velocity of bound binary system used in velocity dispersion calculation

$^{b}$\ not cluster members ($V_{\rm pec} > 3000 $ km~s$^{-1} $)

$^{c}$\ the weak continuum did not allow a good cross-correlation to be carried out

$^{d}$\ the peculiar velocity is calculated from the \emph{emission} velocity
\end{table*}

\subsection{Comparison of emission and absorption redshifts}

Comparing the velocities obtained from emission lines
(Table~\ref{tab:vem}) and absorption features (Table~\ref{tab:vabs}),
it is evident that, although the results are all consistent within the
errors, there is a systematic $V^{\rm(abs)} - V^{\rm(em)} = 65\pm9$
km~s$^{-1}$ (leaving out NGC~6251, which will be considered
separately).  This could be due to the slight defocussing effect
described in section~\ref{sec:redshifts}.  Alternatively it could
indicate that the adopted velocity of the template galaxy NGC~7619 is
a slight over-estimate (see section~\ref{sec:obs}).  Making a
correction of 65 km~s$^{-1}$ to the $V^{\rm(em)}$ of object 8 (the
only \emph{emission} velocity used in the velocity dispersion
calculation) has a negligible effect on the systemic velocity and on
$\sigma_{z}$.

Part of the unusually high discrepancy between the emission and
absorption redshifts for NGC~6251 can be ascribed to this systemic
error.  The remaining $\sim100$ km~s$^{-1}$ difference is in agreement
with the results of Ferrarese \&\ Ford (1999), who also find the
emission lines to be at a significantly lower velocity than the
absorption features.  Adding up the gaussian constituents of their
fits to the H$\alpha$~+~[N~II] system over all their (\arcsd{0}{1})
HST apertures produces an emission line spectrum which is remarkably
similar in shape and position to our nuclear (\arcsd{1}{8} aperture)
spectrum.

\subsection{Comparison of velocities with the literature}

The redshifts of several of the galaxies observed here have also been
measured by other authors. They are compared to our results in
Table~\ref{tab:vcomp}. There is considerable discord among the
velocities, with the most severe disagreement arising when comparing
our values to those published in the 1970s and early 1980s.

Of the five galaxies showing serious discrepancies with published
values, one (source 7) has consistent absorption and emission
redshifts in this paper.  We have a very high-quality spectrum for
object 3, and the cross-correlation with the template galaxy yielded a
high Tonry \&\ Davis `R' factor (an indication of the reliability of
the resulting redshift).  The coordinates given by Richter \&\
Huchtmeier (1991) for object 8 are 2\arcm\ away from the correct
position, and Sargent et al.\,(1982) indicate that they are unsure of
the accuracy of their velocity measurement for source 10; in both of
these cases, we suspect that target misidentification in the earlier
observations may be the reason for the different redshift values.

\begin{table}
\caption{Comparison of measured velocities with the literature}
\label{tab:vcomp}
\begin{tabular}{lr@{\,$\pm$\,}lr@{\,$\pm$\,}ll}
\hline
Object & \multicolumn{2}{c}{This paper} & \multicolumn{2}{c}{Published values} & Reference$^{\,a}$ \\
       & \multicolumn{2}{c}{km s$^{-1}$}& \multicolumn{2}{c}{km s$^{-1}$}      & \\
\hline
1      & 7459           & 22      & 7400      & 22         & Z\emph{cat} 95 \\
\multicolumn{3}{c}{}              & 6900      & 42         & RC3 91         \\
\multicolumn{3}{c}{}              & 7380      & 100        & SRdB 82        \\
\multicolumn{3}{c}{}              & 7294      & 60         & SO 81          \\
\multicolumn{3}{c}{}              & 6900      & 600        & MO 79          \\
3      & 7240           & 46      & 6510      & 100        & SRdB 82        \\
4      & 7001           & 38      & 6840      & 100        & SRdB 82        \\
6      & 7773           & 30 $^{\,b}$ & 7789  & 24         & SHDYFT 92      \\
\multicolumn{3}{c}{}              & 7710      & 100        & SRdB 82        \\
7      & 4224           & 140     & 17010     & 100        & SRdB 82        \\
8      & 7621           & 61      & 1774      & ---        & RH 91          \\
9      & 7554           & 56      & 7320      & 100        & SRdB 82        \\
\multicolumn{3}{c}{}              & 20208     & 184        & GC 73          \\
10     & 16644          & 6       & 7590      & 100 $^{\,c}$ & SRdB 82      \\
11     & 7156           & 57 $^{\,b}$ & 6930  & 100        & SRdB 82        \\
12     & 6548           & 40      & 6395      & 143        & GC 73          \\
\hline
\end{tabular}

$^{a}$\ references:

\begin{tabular}{ll}
Z\emph{cat} 95 & Huchra et al., 1995, CfA Redshift Catalogue,           \\
               & Harvard-Smithsonian Center for Astrophysics            \\
SHDYFT 92      & Strauss et al., 1992, ApJS, 83, 29                     \\
RC3 91         & de Vaucouleurs et al., 1991, Third Reference           \\
               & Catalogue of Bright Galaxies, Springer Verlag:         \\
               & New York                                               \\
RH 91          & Richter \&\ Huchtmeier, 1991, A\&AS, 87, 425           \\
SRdB 82        & Sargent et al., 1982, unpublished                      \\
SO 81          & Shuder \&\ Osterbrock, 1981, ApJ, 250, 55              \\
MO 79          & Miley \&\ Osterbrock, 1979, PASP, 91, 257              \\
GC 73          & Gregory \&\ Connolly, 1973, ApJ, 182, 351              \\
\end{tabular}

$^{b}$\ mean velocity of two core components, as previous authors did
not take account of the double nucleus

$^{c}$\ This value is placed between brackets in the SRdB 82 paper,
indicating that the authors are not entirely confident about it.

\end{table}

\section{Cluster systemic redshift and velocity dispersion}\label{sec:veldisp}

We used the cross-correlation results in Table~\ref{tab:vabs} to
compute the systemic redshift and the velocity dispersion of the
cluster of galaxies using the method of Danese, De Zotti \& di Tullio
(1980) and as described by Werner, Worrall \& Birkinshaw (1999).

The mean velocity of the cluster was found to be $c\overline{z} =
7193\pm 128$ km~s$^{-1}$, equivalent to a redshift $\overline{z} =
0.0240 \pm 0.0004$. The one-dimensional line-of-sight velocity
dispersion $\sigma_{z}= 388\, (+128, -65)$ km~s$^{-1}$, yields a
three-dimensional physical velocity dispersion of $\sigma= 672\,
(+239, -145)$ km~s$^{-1}$.

Such a low velocity dispersion suggests that the cluster is poor.  A
comparison with the correlation between $\sigma$ and Abell richness
class R (e.g. Danese et al.\,1980) indicates $\rm{R} \leq 0$,
corresponding to a population $\leq 30$.

As the results in Table~\ref{tab:vabs} indicate, NGC~6251 has a
peculiar velocity of $260 \pm 130 $ km~s$^{-1} = 0.67 \pm 0.33$
$\sigma_{z}$ within the cluster.  It is generally believed that giant
elliptical radio galaxies dominating clusters are more or less at rest
relative to the cluster's systemic velocity.  Although this is
consistent at the two sigma level with the data above, there is a
reason that the peculiar velocity of NGC~6251 given in
Table~\ref{tab:vabs} may be an overestimate.  Considering the
projected distances of objects 12 and 13 from the rest of the cluster
($>$ 5 Mpc), their comparatively large peculiar velocities ($\geq 1.45
\sigma_{z}$), and their closeness to one another in position and
velocity, it seems likely that they are part of the
physically-distinct IC~1143 group, and that they should be removed
from the velocity dispersion calculations.  This leads to a
significantly higher systemic velocity for the cluster, $c\overline{z}
= 7328\pm 105$ km~s$^{-1}$ ($\overline{z}= 0.0244 \pm 0.0004$) with
$\sigma_{z}= 283\, (+109, -52)$ km~s$^{-1}$.  In this case, NGC~6251
is one of the slowest-moving members of the cluster with a peculiar
velocity of 128 $\pm$ 107 km~s$^{-1}$.

\section{The dynamics of NGC~6251}\label{dynamics}

Heckman et al.\,(1985b) measured the rotation curves of several
radio-loud and radio-quiet elliptical galaxies to test theories that
the rotation axis might coincide with the radio axis in powerful radio
galaxies (log $P_{\rm 178\ MHz} \geq$ 24 W Hz$^{-1}$).  They present
rotation curves for NGC~6251 in position angles of $27^{\circ}$ and
$124^{\circ}$ and conclude that the galaxy rotates about an axis at
P.A. $\sim 85^{\circ} \pm 28^{\circ}$ with a velocity of $V \sim 47
\pm 16$ km s$^{-1}$.

We measured rotation curves for two position angles, 25.3$^{\circ}$
and 115.3$^{\circ}$ (Table~\ref{tab:obsh}), the latter being aligned
with the jet. The results of the slit in P.A. = 25.3$^{\circ}$ suggest
a slow rotation of the order of 50 km s$^{-1}$ about the jet,
consistent with Heckman et al.'s result in P.A. = 27$^{\circ}$.  The
velocities \emph{along} the jet were inconclusive.  Deeper
observations are required to obtain sufficient signal-to-noise in the
outer parts of the galaxy (at a radius $\geq$ 8\arcs\ or $\sim$ 6 kpc)
if more precise information on the rotation axis or velocity than that
found by Heckman et al.\,is to be obtained.

\section{Line emission along the radio jet?}\label{sec:jetem}

The jet of the powerful radio galaxy M87 has been detected at optical
wavelengths, and many authors report imaging and spectroscopic studies
(see Keel 1988 and references therein).  Keel subsequently detected
optical continuum emission from the NGC~6251 jet, and his B-, V- and
R-band fluxes of feature A (10\arcs -- 40\arcs\ or $\sim$ 7 -- 29 kpc
from the nucleus) imply a spectrum much steeper than M87 or other
similar jets.

Our observations were not deep enough to reveal significant continuum
emission from feature A, but we searched for line emission from
[O~III], H$\alpha$ and [N~II].  No emission lines in the jet were
found, at an upper limit (for H$\alpha$) of $10^{-17}-10^{-16}$ erg
cm$^{-2}$ s$^{-1}$ for a range of line widths of $300-1000$ \AA,
respectively.  Poor weather prevented us from obtaining the planned,
much deeper, exposures.

\section{Discussion}\label{sec:disc}

\subsection{NGC 6251 and Zw~1609.0+8212}

Considering our redshifts in the context of the two subsystems that
make up cluster Zw~1609.0+8212 (see section~\ref{sec:intro}), it is
clear that our sample does not include any members of the subcluster
separately listed as A2247 ($c\overline{z} = 11670$ km~s$^{-1}$), and
situated in the Eastern part of the Zwicky cluster, South-East of
NGC~6251.  Objects 12 and 13 appear to belong to a slightly closer
group or small cluster on the Western edge of Zw~1609.0+8212 and we
have accordingly argued for their removal from the velocity dispersion
calculations in section~\ref{sec:veldisp}.  Objects 7 and 10 have
redshifts unlike those of other objects in the sample, and appear to
be more or less isolated field galaxies.  This leaves us with 8
companion galaxies (two with double nuclei) near the velocity of
NGC~6251, and these seem to form a third sub-cluster within the limits
of Zw~1609.0+8212.  It is clear that not all the brightest galaxies
within Zwicky's boundary are physically associated.

\begin{figure*}
\hfill
\psfig{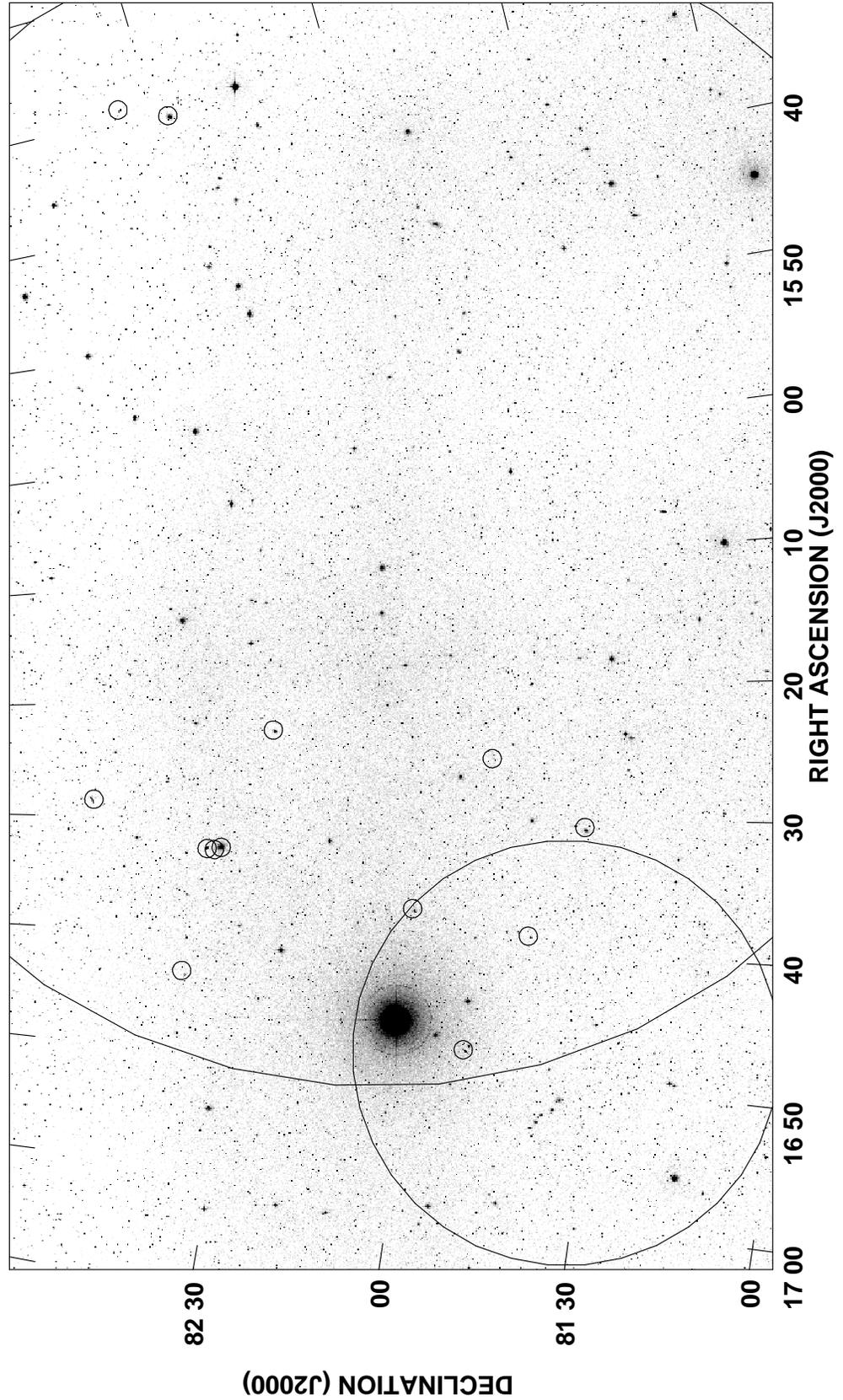}
\begin{minipage}{15 cm}

\caption{Second Digitized Sky Survey (DSS2) image with our targets
indicated by small circles.  The $\sim1^{\circ}- \rm diameter$ circle
in the South-Eastern corner of the field is the outline of Abell 2247.
The large circle indicates the bounds of Zwicky cluster
Zw~1609.0+8212.}

\label{fig:dssfield}
\end{minipage}
\end{figure*}

Figure~\ref{fig:dssfield} shows the positions of our sources, as well
as rough outlines of clusters A2247 and Zw~1609.0+8212.

\subsection{Velocity dispersion and X-ray temperature}

The one-dimensional velocity dispersion of a cluster of galaxies is
related to the X-ray gas temperature of the cluster atmosphere (Lubin
\&\ Bahcall 1993; Bird, Mushotzky \&\ Metzler 1995; Girardi et
al.\,1995).  The higher velocity dispersion reported in
section~\ref{sec:veldisp} corresponds to an X-ray temperature of $T =
1.2 (+0.8, -0.3)\,\rm{keV}$.  The alternative velocity dispersion of
$\sigma_{z}= 283\, (+109, -52)$ km~s$^{-1}$ derived for the cluster
without IC 1139 and IC 1143 (which we argue are part of a separate
group) suggests a lower gas temperature of $T = 0.7 (+0.6,
-0.2)\,\rm{keV}$.

Comparing this temperature of 0.7 keV with the extrapolated $L_{X}-T$
relation of Arnaud \&\ Evrard (1999), we find that it corresponds to a
bolometric X-ray luminosity $L_{X}^{(bol)} \simeq 2.4 (+11.6, -1.5)
\times 10^{42}$ ergs~s$^{-1}$, or $L_{X} \simeq 2.2 (+10.4, -1.4)
\times 10^{42}$ ergs~s$^{-1}$ in the band 0.2--1.9 keV.  Worrall \&\
Birkinshaw (1994) estimate $L_{X}^{(0.2-1.9 \mathrm keV)} = 8 \times
10^{41}$ ergs~s$^{-1}$ for gas within a 3\arcm\ (130 kpc) radius.
Given the parameters of their $\beta$-model fit, this correponds to a
total luminosity of $L_{X}^{(0.2-1.9 \mathrm keV)} \simeq 2 \times
10^{42}$ ergs~s$^{-1}$ for the whole cluster, which is in good
agreement with our velocity dispersion via the $\sigma_{z}-T$ and
$L_{X}-T$ relations.

The predicted temperature for the velocity dispersion also agrees with
two-component spectral fits based on \it ROSAT \rm PSPC data
(Birkinshaw \&\ Worrall 1993), within the rather large uncertainties.
It is clear that the gas that can be confined in this cluster is too
cool for static pressure confinement of the radio jet at distances
between 5 and 200 kpc from the core.

\section{Conclusions}\label{sec:conc}

In this paper we have shown that:

\begin{enumerate}

\item{there is a poor cluster of galaxies associated with NGC~6251; it
is one of at least three distinct and not necessarily related
sub-clusters of Zw~1609.0+8212;}

\item{the temperature of the cluster atmosphere inferred from the
velocity dispersion measurement is not high enough to provide static
pressure confinement of any part of the kpc-scale radio jet (at
distances of 5 to 200 kpc from the core) according to the lower limits
on jet pressure calculated by Birkinshaw \&\ Worrall (1993);}

\item{a tentative rotation curve perpendicular to the direction of the
jet agrees with that measured by Heckman et al.\,(1985b); and}

\item{there is no evidence for [O III], H$\alpha$, or [N II] line
emission at feature A on the jet, where Keel (1988) found optical
continuum emission.}

\end{enumerate}

Better optical spectra of NGC~6251 are needed to probe deeper for line
emission and to confirm the tentative dynamical results presented in
section~\ref{sec:jetem}.  Forthcoming X-ray observations should
measure the temperature of the cluster atmosphere, predicted to be
0.5-2.0 keV, based on our measurements of the cluster velocity
dispersion.

\section*{Acknowledgments}

Observations reported in this paper were obtained at the Multiple
Mirror Telescope Observatory, a facility operated jointly by the
University of Arizona and the Smithsonian Institution.

Second Digitized Sky Survey (DSS2) image obtained from the ESO/ST-ECF
Archive (http://archive.eso.org/dss/dss).

%Digitized Sky Survey data for Northern declinations are based on
%photographic data obtained using the Oschin Schmidt Telescope on Palomar
%Mountain. The Palomar Observatory Sky Survey was funded by the
%National Geographic Society. The Oschin Shmidt Telescope is operated
%by the California Institute of Technology and Palomar Observatory. The
%plates were processed into the present compressed digital format with
%their permission. The Digitized Sky Survey was produced at the Space
%Telescope Science Institute (STScI) under U.S. Goverment grant NAG
%W-2166.

\end{document}